\documentclass[aps,prd,preprintnumbers,superscriptaddress,nofootinbib,floatfix,twocolumn,notitlepage]{revtex4-2}

\usepackage{graphicx}  % needed for figures
\usepackage{dcolumn}   % needed for some tables
\usepackage{bm}        % for math
\usepackage{epsfig,amsmath,amssymb,verbatim,mathrsfs,array,layout,textcomp,amssymb,latexsym,slashed}
\usepackage{xcolor}
\usepackage{comment}

\usepackage[colorlinks=true,citecolor=blue,urlcolor=blue,linktocpage=true,
linkcolor=blue]{hyperref}
\usepackage[utf8]{inputenc}
\usepackage{multirow}
\usepackage{booktabs} %for \toprule
\usepackage{siunitx} %for \num
\usepackage{tabularx} %for tabularx
\usepackage{paralist}
\usepackage{enumitem}
\usepackage{placeins}
\usepackage[normalem]{ulem}
\usepackage{bbold}

\newcommand{\ps}{p\hspace{-0.44em}/\hspace{0.06em}}

\newcommand{\bbms}{\bbs\ mixing}

\newcommand{\bbs}{\ensuremath{B_s\!-\!\Bbar{}_s\,}}

\newcommand{\Bbar}{\,\overline{\!B}}

\def\beq#1\eeq{\begin{align}#1\end{align}}

\newcommand{\bea} {\begin{eqnarray}}
\newcommand{\eea} {\end{eqnarray}}
\newcommand{\ba} {\begin{eqnarray*}}
\newcommand{\ea} {\end{eqnarray*}}
\newcommand{\GeV} {\,\text{GeV}}
\newcommand{\TeV} {\,\text{TeV}}

\newcommand{\hc}{\mathrm{h.c.}}

\newcommand{\eg}{{\em e.g.}}
\newcommand{\ie}{{\em i.e.}}

\usepackage[utf8]{inputenc} % utf8 input good for non-ascii characters
\usepackage[T1]{fontenc} % Good for hyphenating non-ascii characters
\usepackage{lmodern}
\usepackage[british]{babel}
\usepackage{graphicx}
\graphicspath{{figs/}} % Automatically look in the figs directory

\usepackage{microtype} % Improves typesetting
\usepackage{xfrac} % Nice small fractions
\makeatletter
\g@addto@macro\@floatboxreset\centering
\makeatother
\numberwithin{equation}{section} % Reset equation numbers in each section
\allowdisplaybreaks % Allow long maths to be broken over multiple pages

\usepackage[capitalise]{cleveref} % Add automatic reference type text via \cref command
\creflabelformat{equation}{\textup{#2#1#3}} % Removes bracket around equation numbers when referencing

\usepackage{booktabs} % Better tables
\usepackage{tabularx} % Tables with auto width columns
\usepackage{longtable} % Allows tables that span multiple pages
\usepackage{array} % Allows us to define custom column types for tables
\usepackage[italic]{hepnames} % Add particle name macros (e.g. \PBs)
\usepackage{braket} % Adds Dirac bra-ket notation
\usepackage{slashed} % Adds Dirac slash notation
\usepackage{siunitx} % Add support for units
\DeclareSIUnit\barn{b}
\DeclareSIUnit\invfb{\per\femto\barn}
\DeclareSIUnit\invab{\per\atto\barn}
\DeclareSIUnit\invps{\ps^{-1}}
\usepackage{suffix} % Allows us to define a starred command, for CKM macro later

\newcommand{\Lagrangian}{\ensuremath{\mathcal{L}}\xspace}

\newcommand{\V}[1]{\ensuremath{V_{#1}^{}}}
\WithSuffix\newcommand\V*[1]{\ensuremath{V_{#1}^*}}

\newcommand{\CHqone}{C^{(1)}_{Hq}}

\newcommand{\CHqthree}{C^{(3)}_{Hq}}

\newcommand{\CHud}{C_{Hud}}
\newcommand{\CHudbracket}{\left[\CHud\right]}
\newcommand{\program}[1]{\texttt{#1}}
\newcommand{\flavio}{\program{flavio}\xspace}
\newcommand{\smelli}{\program{smelli}\xspace}
\newcommand{\wilson}{\program{wilson}\xspace}
\newcommand{\mmeft}{\program{MatchMakerEFT}\xspace}

\usepackage{xcolor}

\begin{document}
\preprint{PSI-PR-22-09,~ZU-TH-12/22,~KEK-TH-2411}

\title{Large $t\to cZ$ as a Sign of Vector-Like Quarks in Light of the $W$ Mass}

\author{Andreas Crivellin}
\email{andreas.crivellin@cern.ch}
\affiliation{Physik-Institut, Universit\"at Z\"urich, Winterthurerstrasse 190, 8057 Z\"urich, Switzerland}
\affiliation{Paul Scherrer Institut, 5232 Villigen PSI, Switzerland}

\author{Matthew Kirk}
\email{mjkirk@icc.ub.edu}
\affiliation{Departament de F\'isica Quàntica i Astrof\'isica (FQA), Institut de Ciències del Cosmos (ICCUB), Universitat de Barcelona (UB), Spain}

\author{Teppei Kitahara}
\email{teppeik@kmi.nagoya-u.ac.jp}
\affiliation{Institute for Advanced Research, Nagoya University, Nagoya 464--8601, Japan}
\affiliation{Kobayashi-Maskawa Institute for the Origin of Particles and the Universe,
Nagoya University,  Nagoya 464--8602, Japan}
\affiliation{Theory Center, IPNS, High Energy Accelerator Research Organization (KEK), Tsukuba 305--0801, Japan}

\author{Federico Mescia}
\email{mescia@ub.edu}
\affiliation{Departament de F\'isica Quàntica i Astrof\'isica (FQA), Institut de Ciències del Cosmos (ICCUB), Universitat de Barcelona (UB), Spain}

\begin{abstract}
\noindent
The rare flavour changing top quark decay $t\to cZ$ is a clear sign of new physics and experimentally very interesting due to the huge number of top quarks produced at the LHC. However, there are few (viable) models which can generate a sizable branching ratio for $t\to cZ$ -- in fact vector-like quarks seem to be the only realistic option. In this paper, we investigate all three representations (under the Standard Model gauge group) of vector-like quarks ($U$, $Q_1$ and $Q_7$) that can generate a sizable branching ratio for $t\to cZ$ without violating bounds from $B$ physics. Importantly, these are exactly the three vector-like quarks which can lead to a sizable positive shift in the prediction for $W$ mass, via the couplings to the top quark also needed for a sizable Br($t\to cZ$). Calculating and using the one-loop matching of vector-like quarks on the Standard Model Effective Field Theory, we find that Br($t\to cZ$) can be of the order of $10^{-6}$, $10^{-5}$ and $10^{-4}$ for $U$, $Q_1$ and $Q_7$, respectively and that in all three cases the large $W$ mass measurement can be accommodated.
\end{abstract}

\maketitle

\section{Introduction}\label{sec:intro}

The Standard Model (SM) of particle physics contains three generations of chiral fermions, \ie~Dirac fields whose left and right-handed components transform differently under its gauge group. While a combination of LHC searches and flavour observables excludes a chiral $4^{\rm th}$ generation~\cite{Eberhardt:2012ck,Eberhardt:2012gv}, vector-like fermions (VLFs) can be added consistently to the SM without generating gauge anomalies. In fact, VLFs appear in many extensions of the SM such as grand unified theories~\cite{Hewett:1988xc,Langacker:1980js,delAguila:1982fs}, composite models or models with extra dimensions~\cite{Antoniadis:1990ew,Arkani-Hamed:1998cxo} and little Higgs models~\cite{Arkani-Hamed:2002ikv,Han:2003wu} (including the option of top condensation~\cite{Dobrescu:1997nm,Chivukula:1998wd,He:2001fz,Hill:2002ap,Anastasiou:2009rv}).

VLFs are not only interesting from the theoretical perspective, but also from the phenomenological point of view as they could be involved in an explanation of $b\to s\ell^+\ell^-$ data~\cite{Altmannshofer:2014cfa,Gripaios:2015gra,Arnan:2016cpy,Arnan:2019uhr,Crivellin:2020oup}, the tension in $(g-2)_\mu$~\cite{Czarnecki:2001pv,Kannike:2011ng,Dermisek:2013gta,Freitas:2014pua,Belanger:2015nma,Aboubrahim:2016xuz,Kowalska:2017iqv,Raby:2017igl,Choudhury:2017fuu,Kowalska:2017iqv,Calibbi:2018rzv,Capdevilla:2020qel,Capdevilla:2021rwo,Crivellin:2021rbq,Calibbi:2021pyh,Arcadi:2021glq, Paradisi:2022vqp} or account for the Cabibbo angle anomaly~\cite{Belfatto:2019swo,Grossman:2019bzp,Seng:2020wjq,Coutinho:2019aiy,Crivellin:2020lzu,Endo:2020tkb,Kirk:2020wdk,Belfatto:2021jhf,Branco:2021vhs,Balaji:2021lpr}. Furthermore, vector-like quarks (VLQs) can lead to tree-level effects in $Z$-$t$-$c$ and $h$-$t$-$c$ couplings after electroweak (EW) symmetry breaking, and therefore generate sizeable effects in the related flavour-changing neutral current (FCNC) decays of the top quark~\cite{Nardi:1995fq, Cacciapaglia:2011fx,Botella:2012ju,Okada:2012gy,Belfatto:2021jhf,Branco:2021vhs,Balaji:2021lpr}.

There are three VLQs ($U$, $Q_1$ and $Q_7$) that generate a $Z$-$t$-$c$ (and $h$-$t$-$c$) coupling but do not give rise to down-quark FCNCs at tree-level, such that the former can be sizable. However, even these VLQs affect e.g.~the $W$ mass
\footnote{
The contribution of VLQs to the $W$ mass, via the oblique $S$ and $T$ parameters, has previously been calculated at fixed order in Ref.~\cite{Chen:2017hak}, where they studied the contribution to electroweak observables and Higgs decays only.
}
and $B$ decays at the loop-level. Therefore, it is important to calculate and include these effects in a phenomenological analysis in order to assess the possible size of $t\to Z(h)c$ and to evaluate if one can account for the recent measurement of the $W$ mass by the CDF collaboration~\cite{CDF:2022hxs}, which suggests that $M_W$ is larger than the expected within the SM.

\section{Setup and Matching Calculation}
\label{sec:generic}

There are seven possible representations (under the SM gauge group $SU(3)_C\times SU(2)_L\times U(1)_Y$) of VLQs, given in Table~\ref{VLQrepresentations}, defining them as heavy fermions which are triplets of $SU(3)_C$ and that can mix with the SM quarks after EW symmetry breaking, \ie~fermions which can have couplings to the SM Higgs and a SM quark. The kinetic and mass terms\footnote{Note that mass terms such as $m_{i}^U \bar U {u_i}$ can always be removed by a field redefinition, such that the kinetic terms and the mass terms take the diagonal form shown in Eq.~\eqref{eq:VLQ_lagrangian}.}
 are
\begin{equation}
\label{eq:VLQ_lagrangian}
\mathcal{L} = \sum_{F} \bar{F} \left( i  \slashed{D}  - M_F \right) F\,,
\end{equation}
where $F=\{U,\,D,\,Q_1,\,Q_5,\,Q_7,\,T_1,\,T_2\}$ and
\begin{equation}
   D_{\mu} = \partial_{\mu}
   + i g_1 Y_F B_{\mu}
   + i g_2 S^I W^I_{\mu}
   + i g_s T^A G^{A}_{\mu}\,.
   \label{covariantD}
\end{equation}
Here $T^A = \frac{1}{2}\lambda^A$ and $(S^I)_{jk}$ are $0$, $\frac{1}{2}(\tau^I)_{jk}$, and $-i \epsilon_{Ijk}$ for the $SU(2)_L$ singlet, doublet, and triplet representations, respectively, and $\lambda^A$ and $\tau^I$ are the Gell-Mann and the Pauli matrices. The (generalized) Yukawa couplings are encoded in the Lagrangian
\begin{align}
{{\cal L}_{}} =& {{\cal L}^H_{qq}}
+ {{\cal L}^H_{q \text{VLQ}}}
+ {{\cal L}^H_{\text{VLQ} \text{VLQ}}} \,,
\end{align}
where the first term contains the SM Yukawa couplings
\begin{align}
\begin{aligned}
 -{{\cal L}^H_{qq}} =&\, Y^u_{ij} \bar{q}_i \tilde{H} u_j + Y^d_{ij} \bar{q}_i H d_j + \text{h.c.}\,,
 \end{aligned}
\end{align}
the second term the Higgs interactions with vector-like and SM quarks
\begin{align}
- {{\cal L}^H_{q \text{VLQ}}} &=
\xi _{i}^U{{\bar U}}{{\tilde H}^\dag }{q_i} + \xi _{i}^D{{\bar D} }{H^\dag }{q_i} +
\xi _{i}^{u_1}{{\bar Q}_{1} }\tilde H{u_i} \nonumber\\
&+ \xi _{i}^{d_1}{{\bar Q}_{1} }H{d_i} +  \xi _{i}^{{Q_5}}{{\bar Q}_{5}}\tilde H{d_i} + \xi _{i}^{{Q_7}}{{\bar Q}_{7}}H{u_i} \\
&+\frac{1}{2}\xi _{i}^{{T_1}}{H^\dag }\tau \cdot{{\bar T}_{1}}{q_i} + \frac{1}{2}\xi _{i}^{{T_2}}{{\tilde H}^\dag }\tau \cdot{{\bar T}_{2}}{q_i} + \hc \,,\nonumber
\end{align}
and the last term defines the Higgs interactions with two VLQs (given in the supplementary material  as they are not relevant for our analysis). Here $i,j = \{1,2,3\}$ are flavour indices and $\tau\cdot\bar T=\sum_I \tau^I \bar T^I$.

\begin{table}[t]
\centering
\renewcommand{\arraystretch}{1.2}
  \scalebox{1}{
\begin{tabular}[t]{l |  c c c c | c c c c c c c }
\hline
&$u$&$d$&$q$&$H$&\bf{$U$}&$D$&\bf{$Q_1$}&$Q_5$&\bf{$Q_7$}&$T_1$&$T_2$\\
\hline
$SU(3)_C$&3&3&3&1&3&3&3&3&3&3&3\\
$SU(2)_L$&1&1&2&2&1&1&2&2&2&3&3\\
% $U(1)_Y$&2/3&$-$1/3&1/6&1/2&2/3&$-$1/3&1/6&$-$5/6&7/6&$-$1/3&2/3\\
$U(1)_Y$&\sfrac{2}{3}&$-$\sfrac{1}{3}&\sfrac{1}{6}&\sfrac{1}{2}&\sfrac{2}{3}&$-$\sfrac{1}{3}&\sfrac{1}{6}&$-$\sfrac{5}{6}&\sfrac{7}{6}&$-$\sfrac{1}{3}&\sfrac{2}{3}\\
\hline
\end{tabular}
}
\caption{Representations of the Higgs, the SM quarks and of the VLQs under the SM gauge group. The three representations in bold are the ones relevant for our analysis as they generate flavour-changing top decays at tree level but down-quark FCNCs first appear at one-loop level.}
\label{VLQrepresentations}
\end{table}

\subsection{SMEFT and Matching}
\begin{figure}[t]
\hspace{-0.1cm}
\includegraphics[width=0.249\textwidth]{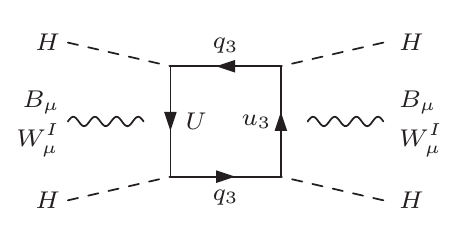}
\includegraphics[width=0.228\textwidth]{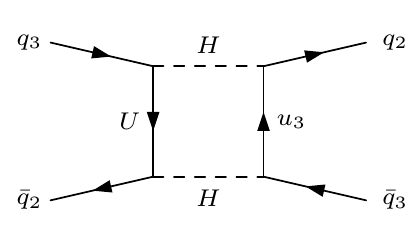}
\caption{
Examples of Feynman diagrams showing the $U$ contributions to the operator $Q_{HD}$, affecting the $W$-boson mass  (left), and $Q_{qq}^{(1,3)}$, affecting \bbms~(right).}
\label{fig:QHDU}
\end{figure}

We write the SMEFT Lagrangian as
\begin{equation}
\Lagrangian_\text{SMEFT} = \mathcal{L}_{\rm SM} + \sum_i C_i Q_i\,,
\end{equation}
such that the Wilson coefficients have dimensions of inverse mass squared. Using the Warsaw basis~\cite{Grzadkowski:2010es}, the operators generating modified gauge-boson couplings to quarks are
\begin{align}
Q_{H q}^{(1)} \,, Q_{H q}^{(3)} \,, Q_{H u} \,, Q_{H d} \,, Q_{H ud}\,,
\end{align}
and the four-quark operators generating $\Delta F=2$ processes read
\begin{align}
&Q^{(1)}_{qq}\,,
Q^{(3)}_{qq} \,,
Q_{uu}  \,,
Q_{dd} \,,
% Q^{(1)}_{ud}  \,,
% Q^{(8)}_{ud} \,,
% \nonumber
% \\
% &
Q^{(1)}_{qu} \,,
Q^{(1)}_{qd}  \,,
Q^{(8)}_{qu} \,,
Q^{(8)}_{qd} \,,
% Q^{(1)}_{quqd} \,,
% Q^{(8)}_{quqd}  \,.
\end{align}
The explicit definitions of all these operators can be found in Ref.~\cite{Grzadkowski:2010es} and in the supplementary material. The dipole operators, responsible for radiative down-type quark decays after EW symmetry breaking, are $Q_{dW}$ and $Q_{dB}$. In addition, we have the operator involving three Higgs fields,
$Q_{uH}$, that generates modifications of the Higgs-up-quark coupling, including possibly flavour changing ones, after EW symmetry breaking. Finally we also need two bosonic operators that lead to a modification to the $W$ mass, $Q_{HD}$ and $Q_{HWB}$, with their contributions approximately given by
\begin{equation}
\label{eq:delta_MW}
\delta M_W \approx -v^2 (29\, C_{HD} + 64\, C_{HWB} + \cdots) \, \mathrm{GeV} \,,
\end{equation}
where $v\simeq 246$\,GeV and ($\cdots$) indicates SMEFT operators not relevant in our scenario with VLQs.\footnote{
Note that the SMEFT effects in the $W$ mass are known fully at leading order~\cite{Berthier:2015oma,Bjorn:2016zlr}, but only partially at next-to-leading order (NLO)~\cite{Dawson:2019clf}, since in that work flavour universality of the SMEFT coefficients is assumed.
However we have checked that, after making some conservative assumptions about the flavour dependence, the NLO effects are small.}
An example diagram for the $W$ mass correction is shown on the left in Fig.~\ref{fig:QHDU}.

The tree-level matching of the operators generating modified $Z$-quark couplings is given by
\begin{align}
\label{eq:VLQ_SMEFT_matching}
C_{H q}^{(1)ij} + C_{H q}^{(3)ij} & = - \frac{{\xi _{i}^{D*}\xi _{j}^D}}{{2M_{{D }}^2}}
- \frac{{\xi _{i}^{{T_1}*}\xi _{j}^{{T_1}}}}{{8M_{{T_{1 }}}^2}}
+ \frac{{\xi _{i}^{{T_2}*}\xi _{j}^{{T_2}}}}{{4M_{{T_{2 }}}^2}}
\,,\nonumber\\
C_{H q}^{(1)ij} - C_{H q}^{(3)ij} & =
\frac{{\xi _{i}^{U*}\xi _{j}^U}}{{2M_{{U }}^2}}
- \frac{{\xi _{i}^{{T_1}*}\xi _{j}^{{T_1}}}}{{4M_{{T_{1 }}}^2}}
+ \frac{{\xi _{i}^{{T_2}*}\xi _{j}^{{T_2}}}}{{8M_{{T_{2 }}}^2}}
\,,\nonumber\\
C_{H u}^{ij} &=
- \frac{{\xi _{i}^{u_1*}\xi _{j}^{u_1}}}{{2M_{{Q_1 }}^2}}
+ \frac{{\xi _{i}^{{Q_7}*}\xi _{j}^{{Q_7}}}}{{2M_{{Q_{7}}}^2}}
\,,\\
C_{H d}^{ij} &=
\frac{{\xi _{i}^{d_1*}\xi _{j}^{d_1}}}{{2M_{{Q_1 }}^2}}
- \frac{{\xi _{i}^{{Q_5}*}\xi _{j}^{{Q_5}}}}{{2M_{{Q_{5}}}^2}}
\,,\nonumber
\end{align}
for $Z$-$d^i_L$-$d^j_L$, $Z$-$u^i_L$-$u^j_L$, $Z$-$u^i_R$-$u^j_R$, and $Z$-$d^i_R$-$d^j_R$ respectively.
Modified $W$ couplings to left-handed quarks arise from $C_{Hq}^{(3)}$ alone, while right-handed modifications do not appear in our scenario, due to our (later) choice to set $\xi^{d_1}$ to zero which removes all contributions to the $C_{Hud}$ coefficient.
From these equations, we can see that only the representations $U$, $Q_1$ with coupling $\xi^{u_1}$ and $Q_7$ (shown in bold in Table~\ref{VLQrepresentations}) lead to effects in $t\to cZ$ while avoiding tree-level FCNCs in the down sector. An approximate formula for this branching ratio is
\begin{equation}
\!\!\!\! \text{Br}(t \to c Z) \approx \frac{v^4}{2} \left\{  \left[ C_{H q}^{(1)23} \!-\! C_{H q}^{(3)23} \right]^2 + \left[C_{H u}^{23} \right]^2 \right\}\,.
\end{equation}

We calculated the one-loop matching on the SMEFT for these VLQs for the operators relevant for $B$ physics, the $W$ mass and EW precision observables (EWPOs) using \href{https://ftae.ugr.es/matchmakereft}{\mmeft}~\cite{Carmona:2021xtq} and compared the results to our own calculation, finding perfect agreement. Details of our calculation and explicit expressions for the relevant Wilson coefficients are given in the supplementary material.

\section{Phenomenological analysis}
\label{sec:pheno}

The current 95\%~CL upper bounds for $t\to cZ$ and $t\to ch$, based on the full LHC Run 2 data set,
are~\cite{ATLAS:2018zsq,ATLAS:2018jqi,ATLAS:2021stq,VelosoMoriond2022}
\begin{equation}
{\rm Br}(t\to cZ)<1.3\times 10^{-4}\,,
\quad
{\rm Br}(t\to ch)<9.9\times 10^{-4}\,.
\end{equation}
While this already constrains some beyond the SM scenarios, at the high-luminosity (HL-)LHC~\cite{Liss:2013hbb,ATLAS:2019pcn}, FCC-hh~\cite{FCC:2018vvp}, ILC~\cite{ILCInternationalDevelopmentTeam:2022izu}, or the FCC-ee~\cite{Khanpour:2014xla}, one can expect to be sensitive to $t \to c Z$ branching ratios on the order of $10^{-5}$ to $10^{-6}$ \cite{Liu:2020bem,ILCInternationalDevelopmentTeam:2022izu}. For $t\to ch$, see Ref.~\cite{Liu:2020kxt} and references therein, sensitivities on the order of $10^{-4}$ and $10^{-5}$ for the HL-LHC~\cite{TheATLAScollaboration:2013nbo} and FCC-hh~\cite{Liu:2020kxt,Khanpour:2019qnw,Papaefstathiou:2017xuv} are estimated, respectively.
A summary of the future prospects for these FCNC top decays is given in \cref{tab:tcZh_exp_summary}.

\begin{table}
\centering
\begin{tabular}{lll}
\toprule
 & $\mathrm{Br}(t \to c Z) \times 10^5$ & $\mathrm{Br}(t \to c h) \times 10^5$ \\
\midrule
Current LHC & \multirow{2}{*}{\num{13} \cite{ATLAS:2021stq}} & \multirow{2}{*}{\num{99} \cite{VelosoMoriond2022}} \\
($\SI{13}{\TeV}, \SI{139}{\invfb}$) & & \\
\addlinespace
HL-LHC & \num{3.13} \cite{Liu:2020bem} (0\%) & \num{15} \cite{TheATLAScollaboration:2013nbo} \\
($\SI{14}{\TeV}, \SI{3}{\invab}$) & \num{6.65} \cite{Liu:2020bem} (10\%) & \\
\addlinespace
HE-LHC & \num{0.522} \cite{Liu:2020bem} (0\%) & \num{7.7} \cite{Liu:2020kxt} (0\%) \\
($\SI{27}{\TeV}, \SI{15}{\invab}$) & \num{3.84} \cite{Liu:2020bem} (10\%) & \num{8.5} \cite{Liu:2020kxt} (10\%) \\
\addlinespace
FCC-hh & & \multirow{2}{*}{\num{7.7} \cite{Mandrik:2018yhe}} \\
($\SI{100}{\TeV}, \SI{3}{\invab}$) & & \\
\addlinespace
FCC-hh & & \num{2.39} \cite{Papaefstathiou:2017xuv} (5\%) \\
($\SI{100}{\TeV}, \SI{10}{\invab}$) & & \num{9.68} \cite{Khanpour:2019qnw} (10\%) \\
\addlinespace
FCC-hh & \num{0.0887} \cite{Liu:2020bem} (0\%) & \num{0.96} \cite{Liu:2020kxt} (0\%) \\
($\SI{100}{\TeV}, \SI{30}{\invab}$) & \num{3.54} \cite{Liu:2020bem} (10\%) & \num{3.0} \cite{Liu:2020kxt} (10\%) \\
& & \num{4.3} \cite{Mandrik:2018yhe} \\
\addlinespace
ILC & \multirow{2}{*}{\num{9.1} \cite{ILCInternationalDevelopmentTeam:2022izu}} & \\
($\SI{250}{\GeV}, \SI{2}{\invab}$) & & \\
\addlinespace
ILC & \multirow{2}{*}{\num{2.9} \cite{ILCInternationalDevelopmentTeam:2022izu}} & \\
($\SI{1}{\TeV}, \SI{8}{\invab}$) & & \\
\addlinespace
FCC-ee & \multirow{2}{*}{\num{2.8} \cite{Khanpour:2014xla}} & \\
($\SI{350}{\GeV}, \SI{10}{\invab}$) & & \\
\bottomrule
\end{tabular}
\caption{Summary of current limits and future sensitivities for $t\to Z c$ and $t\to h c$. The values in brackets are the assumed systematic uncertainties on the underlying experimental measurements at the future colliders (if provided).}
\label{tab:tcZh_exp_summary}
\end{table}

For the numerical analysis we use the software package \smelli~\cite{Aebischer:2018iyb,Stangl:2020lbh} (based on \flavio~\cite{Straub:2018kue} and \wilson~\cite{Aebischer:2018bkb}),
with $\{\alpha, M_Z, G_F\}$ constituting the input scheme. Furthermore, we work in the down-basis such that Cabibbo-Kobayashi-Maskawa (CKM) elements appear in transitions involving left-handed up-type quarks after EW symmetry breaking, meaning that $Y^d$ is diagonal in unbroken $SU(2)_L$ while $Y^u \approx V^\dagger \cdot \text{diag} (0, 0, y_t)$, with $V$ being the CKM matrix. Note that in our setup the determination of CKM elements is already modified at tree-level. The resulting effects are consistently accounted for in \smelli using the method described in Ref.~\cite{Descotes-Genon:2018foz}, but choosing $\Gamma(K^+ \to \mu^+ \nu) / \Gamma(\pi^+ \to \mu^+ \nu)$, $\mathrm{Br}(B \to X_c e^+ \nu)$, $\mathrm{Br}(B^+ \to \tau^+ \nu)$, and $\Delta M_d / \Delta M_s$ as observables (see supplementary material for details).

Concerning the EW fit, the long standing tension in the $W$ mass, previously with a significance of $\approx$\SI{1.8}{\sigma}~\cite{Zyla:2020zbs,Awramik:2003rn,deBlas:2021wap}, was recently increased by the measurement of the CDF collaboration~\cite{CDF:2022hxs}.
%By combining this with all existing measurements from the Tevatron~\cite{CDF:2022hxs}, LEP~\cite{ALEPH:2010aa} with the LHC ones (ATLAS~\cite{ATLAS:2017rzl} and LHCb~\cite{LHCb:2021bjt}),
%the new world average is~\cite{deBlas:2022hdk}
In~\cite{deBlas:2022hdk}, they have made a naive combination of the existing measurements (Tevatron~\cite{CDF:2022hxs}, LEP~\cite{ALEPH:2010aa}, ATLAS~\cite{ATLAS:2017rzl} and LHCb~\cite{LHCb:2021bjt}), assuming a common \SI{4.7}{\MeV} systematic uncertainty, and give a new world average of
\begin{align}
\label{eq:MW_exp_average}
    M_W^{\rm exp} = 80413.3 \pm 8.0  \,\text{MeV}\,.
\end{align}
This value is \SI{5.5}{\sigma} higher than the SM prediction $M_W^{\rm SM} =  80358.7 \pm 6.0$ {MeV}~\cite{Awramik:2003rn}.

Concerning $B$ physics, even though the hints for lepton flavour universality (LFU) violation in $b\to s\ell^+\ell^-$ data cannot be explained by our LFU effects, an additional LFU part~\cite{Geng:2017svp,Crivellin:2018yvo,Alguero:2018nvb,Alguero:2019ptt,Altmannshofer:2021qrr,Alguero:2021anc,London:2021lfn}, generated by $Z$-$b$-$s$ penguins, can further increase the agreement with data. In addition, box diagrams, like the one shown on the right in Fig.~\ref{fig:QHDU} also generate effect in \bbms ~(we use inputs from Ref.~\cite{DiLuzio:2019jyq} for the SM prediction).

In all our analyses, we set the masses of the VLQs to \SI{2}{\TeV}. This is consistent the published model-independent bounds for third generation VLQs of $M_\text{VLQ} > \SI{1.31}{\TeV}$ limits from ATLAS~\cite{ATLAS:2018ziw} and recent conference reports~\cite{ATLAS:2021ibc,ATLAS:2021ddx} which give slightly stronger limits.
We also checked single VLQ production, which is model-dependent, and found the bounds for our scenarios to be weaker or non-existent.
Let us now consider the three cases of $U$, $Q_1$ and $Q_7$ numerically:

\begin{figure*}[t]
\includegraphics[width=0.75\textwidth]{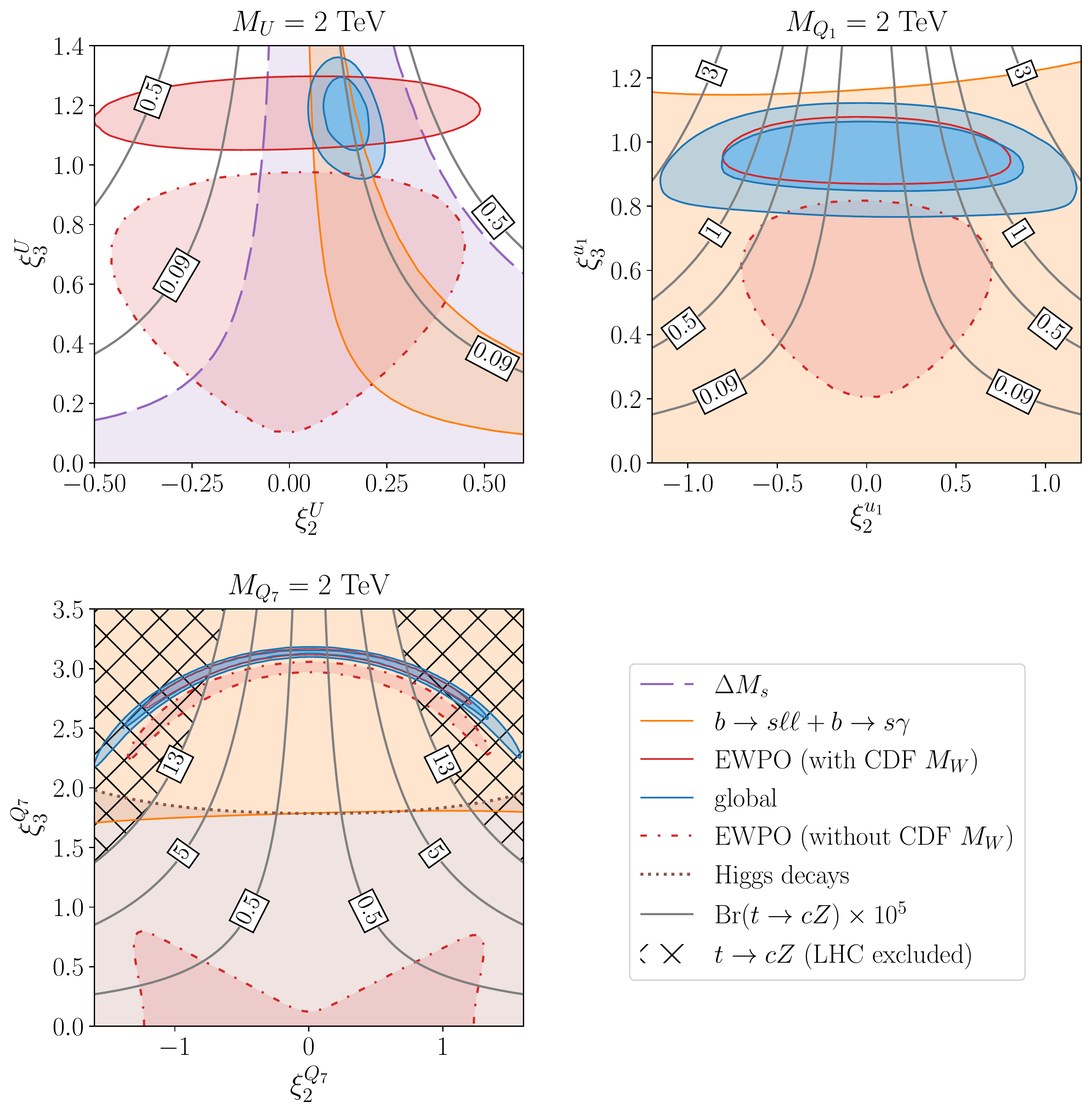}
\caption{
Preferred regions in the $\xi_2$-$\xi_3$ plane for the three representations of VLQ that generate $t\to cZ$ at tree-level but give rise to down-quark FCNCs only at the loop level: $U$ (top-left), $Q_1$ (top-right), and $Q_7$ (bottom-left). The contour lines show the predicted size of $\mathrm{Br}(t\to c Z) \times 10^{5}$. The region preferred by all data (the global fit region with using the new experimental average in Eq.~\eqref{eq:MW_exp_average}) is shown at the \SI{1}{\sigma} and \SI{2}{\sigma} level, while the others regions correspond to \SI{1}{\sigma}. We also show in the preferred region from the EW fit without the inclusion of the new $M_W$ result from CDF (red, dashed-dotted), where it can be seen that a large $t \to cZ$ branching ratio is also possible in this scenario.
Note that in the plot for $Q_7$ the hatched regions on the top-left and top-right are already excluded by the current LHC limits on $t\to cZ$.}
\label{fig:all_fits}
\end{figure*}

{\bf$U$:} In addition to the modified $Z$-$t$-$c$ coupling, this VLQ also generates relevant effects in $b\to s\ell^+\ell^-$ transitions via a $Z$ penguin, resulting in an $C_9 \approx -C_{10}/4$ pattern. In fact, mainly due to the measurements of $P_5^\prime$~\cite{LHCb:2020lmf} and $B_s\to\phi\mu^+\mu^-$~\cite{LHCb:2021xxq,LHCb:2021zwz} there is a preference for a non-zero contribution with such a structure. The bounds from \bbms~turn out to be weakened due to a partial (accidental) cancellation between the one-loop matching and the renormalization group equation (RGE) effect. Similarly, the contribution to $b \to s \gamma$ suffers from a cancellation, but here between terms generated by the matching on the SMEFT and integrating out the $W$ at the weak scale ($b \to s \gamma$ is included within the $b\to s\ell^+\ell^-$ region in \cref{fig:all_fits}). Concerning EWPOs, a shift in $M_W$ is dominantly generated by top-loop effects within the SMEFT (left diagram in Fig.~\ref{fig:QHDU}), bringing theory and experiment into total agreement. Meanwhile, the second generation coupling $\xi^U_2$ is constrained by the total $Z$ width. These finding are summarised in \cref{fig:all_fits} (top-left) where one can see that $\mathrm{Br}(t\to c Z)$ can be of the order of $2\times 10^{-6}$, which could be probed by FCC-hh.

{\bf$Q_1$ with $\xi^{u_1}$:} The VLQ $Q_1$ with the couplings $\xi^{u_1}$ is found to be a very promising candidate for sizable rates of $t \to c Z$, since it has small effects in $B$ physics as it generates at tree-level only right-handed corrections to $Z$-up-quark couplings. At the same time, we can get an improvement concerning the agreement between theory and experiment in $M_W$ through the direct 1-loop contribution to $C_{HD}$ for large couplings is induced through top loops in the SMEFT (thus favouring the third generation coupling), while large couplings to charm quarks are ruled out by the total $Z$ width, as shown in \cref{fig:all_fits} (top-right). From there we see that an enhancement of Br$(t\to cZ)$ up to \num{1e-5} is possible, which could already be probed by the HE-LHC (albeit in an optimistic scenario with zero systematic errors). Note, however, that even in this quite unconstrained scenario $\mathrm{Br}(t \to ch)$ can be at most \num{3e-6}, which is still a factor of three smaller than the reach of even the most optimistic FCC-hh scenario.

{\bf $Q_7$:} In case of the VLQ $Q_7$ (see \cref{fig:all_fits} (bottom-left)), the preferred sign for the contribution in $b\to s\ell^+\ell^-$ processes is generated, but in order for its size to be relevant, quite large couplings are required. Furthermore, for small third generation couplings ($\xi^{Q_7}_3 < 1$) an effect with the wrong sign arises in $M_W$, while for large couplings the sign reverses, which can be traced back to two different contribution, one proportional to $(\xi^{Q_7}_3)^4$ the other involving $(\xi^{Q_7}_3)^2 y_t^2$. Note that in the regime of such large couplings, small tensions with Higgs data arise in the $h \to ZZ, \, W W, \, \gamma \gamma$ partial widths, with tensions of $1.8$, $1.5$, and \SI{1.2}{\sigma}, respectively. Concerning $\mathrm{Br}(t \to c Z)$, again an enhancement of the branching ratio up to \num{1e-5} is possible, which could be probed by the HE-LHC, FCC-hh, FCC-ee, or ILC.
Given the large couplings allowed by data, $\mathrm{Br}(t \to c h)$ can be enhanced up to \num{3e-5}, therefore potentially visible at the FCC-hh if the systematic uncertainties are well controlled.

\section{Conclusions}
\label{sec:conclu}

In this paper we examined the possibility of obtaining a sizable branching ratio for $t\to cZ$ within models containing VLQs. This is only feasible for representations which solely change $Z$ couplings to the up-type quarks at tree-level while not not generating down-type FCNCs at this perturbative order, \ie~$U$, $Q_1$ and $Q_7$. However, at the loop-level, $B$ physics and electroweak observables are still affected. We therefore calculated the one-loop matching of these VLQs onto the SMEFT operators relevant for flavour and electroweak precision observables.

Using these results, we found in our phenomenological analysis that one can generate a sizable branching ratio for $t\to cZ$ of the order of \num{1e-6}, \num{1e-5} and \num{1e-4}, for $U$, $Q_1$ and $Q_7$, respectively. Therefore, the parameter space of $Q_7$ is already constrained by LHC limits on $t\to cZ$, while $Q_1$ and $U$ can be tested by the HL-LHC and the FCC-hh respectively. Importantly, these three VLQ representations are also the ones which lead to a relevant and positive shift in the $W$ mass and can thus explain the larger value of $M_W$, compared to the SM prediction, obtained recently by the CDF collaboration. In fact, accounting for a larger $M_W$ requires sizable couplings to top quarks (see also Ref.~\cite{Heckman:2022the}) which are also important for measurable effects in $t\to cZ$, showing that these observables are correlated. Furthermore, $U$ and $Q_7$ lead to LFU effects in $b\to s\ell^+\ell^-$ which cannot explain $R(K^{(*)})$ but affect observables like $P_5^\prime$ and $B_s\to \phi\mu^+\mu^-$ and, in combination with LFU violating effects, can further improve the description of data. In conclusion, $t\to cZ$ is an unambiguous signal of VLQs and sizable branching ratios of it, within the range of the HL-LHC, are motivated by the recent CDF measurement of the $W$ mass.

\begin{acknowledgments}
A.\ C.~gratefully acknowledges financial support by the Swiss National Science Foundation (PP00P\_2176884).
T.\ K.~is supported by the Grant-in-Aid for Early-Career Scientists (No.\,19K14706) and by the JSPS Core-to-Core  Program (Grant No.\,JPJSCCA20200002) from the Ministry of Education, Culture, Sports, Science, and Technology (MEXT), Japan.
M.\ K. and F.\ M. acknowledge financial support from the State Agency for Research of the Spanish Ministry of Science and Innovation through the “Unit of Excellence Mar\'ia de Maeztu 2020-2023” award to the Institute of Cosmos Sciences (CEX2019-000918-M) and from PID2019-105614GB-C21 and 2017-SGR-929 grants.
\end{acknowledgments}

\bibliographystyle{utphys28mod}
\bibliography{BIB}

%\clearpage
\onecolumngrid
%\appendix

\setcounter{equation}{0}
\setcounter{figure}{0}
\setcounter{table}{0}
\setcounter{section}{0}
\makeatletter
\renewcommand{\theequation}{S.\arabic{section}.\arabic{equation}}
\renewcommand{\thefigure}{S.\arabic{figure}}
\renewcommand{\thetable}{S.\arabic{table}}
\renewcommand{\thesection}{\arabic{section}}

%H\newpage
%%%%%%%%%%%%%%%%%%%%%%%%%%%%%%%%%%%%%%%%%%%%%%%%%%%%%%%%%%%%%%%%%%%%%%
\begin{center}
 \large{\bf Supplemental Material}
 \end{center}
% \begin{center}
% \vspace{10px}
% \large{\bf Supplemental Material}
% \end{center}
%%%%%%%%%%%%%%%%%%%%%%%%%%%%%%%%%%%%%%%%%%%%%%%%%%%%%%%%%%%%%%%%%%%%%%

\section{SMEFT operators}

Here we give the definitions of the SMEFT operators relevant for flavour and electroweak precision observables according to Ref.~\cite{Grzadkowski:2010es}. $p$, $r$, $s$ and $t$ are flavour indices, while color as well as $SU(2)_L$ indices are contracted within the bi-linears and $T^A$ stands for the generators of $SU(3)_C$.

\paragraph*{Modified gauge boson couplings:}
The operators generating modified gauge boson couplings to quarks after EW symmetry breaking are
\begin{align}
Q_{H q}^{(1)ij}&= (H^{\dagger}i\overleftrightarrow{D_{\mu}}H)(\bar{q}_i\gamma^{\mu}q_j)\,,
&
Q_{H q}^{(3)ij} &= (H^{\dagger}i\overleftrightarrow{D_{\mu}^I}H)(\bar{q}_i\tau^I\gamma^{\mu}q_j)\,,
\\
Q_{H u}^{ij} &= (H^{\dagger}i\overleftrightarrow{D_{\mu}}H)(\bar{u}_i\gamma^{\mu}u_j)\,,
&
Q_{H d}^{ij} &= (H^{\dagger}i\overleftrightarrow{D_{\mu}}H)(\bar{d}_i\gamma^{\mu}d_j)\,,
\\
Q_{H ud}^{ij} &= i(\tilde{H}^{\dagger}D_{\mu}H)(\bar{u}_i\gamma^{\mu}d_j)\,,
\label{eq:HiggsqOperators}
\end{align}
with the covariant derivative given in Eq.~(II.2) and
\begin{equation}
\overleftrightarrow{D_{\mu}} = (D_{\mu} - \overset{\leftarrow}{D}_{\mu})\,,\qquad \overleftrightarrow{D_{\mu}^I} = (\tau^I D_{\mu} - \overset{\leftarrow}{D}_{\mu}\tau^I)\,.
\end{equation}

It is useful to write explicitly the modifications of the \PW and \PZ couplings (after EW symmetry breaking) as a function of the SMEFT coefficients:
\begin{align}
\label{eq:ZW_modifications}
\delta \Lagrangian_{W,Z} =
&- v^2 \frac{g}{\sqrt{2}} W^+_\mu \, \bar{u}_i \gamma^\mu \left( \left[ V \cdot \CHqthree \right]_{ij} P_L + \frac{1}{2} \CHudbracket_{ij} P_R \right) d_j + \hc \\
&- v^2 \frac{g}{2 c_W} Z_\mu \, \bar{u}_i \gamma^\mu \left( \left[ V \cdot \left\{ \CHqthree - \CHqone \right\} \cdot V^\dagger \right]_{ij} P_L - \left[C_{Hu}\right]_{ij} P_R \right) u_j \\
&- v^2 \frac{g}{2 c_W} Z_\mu \, \bar{d}_i \gamma^\mu \left( \left[ \CHqthree + \CHqone \right]_{ij} P_L + \left[C_{Hd}\right]_{ij} P_R \right) d_j \,,
\end{align}
where $V$ is the CKM matrix and $v\simeq 246$\,GeV.

\paragraph*{$\Delta F=2$ processes:}
The operators giving rise to, \eg, \bbms, read
\begin{align}
\left[ Q^{(1)}_{qq} \right]_{prst} &= (\bar{q}_p \gamma^\mu q_r)(\bar{q}_r \gamma_\mu q_t)
\,,
&
 \left[ Q^{(3)}_{qq} \right]_{prst} &= (\bar{q}_p \gamma^\mu \tau^I q_r)(\bar{q}_r \gamma_\mu \tau^I q_t) \,,
\\
\left[ Q_{uu} \right]_{prst} &= (\bar{u}_p \gamma^\mu u_r)(\bar{u}_s \gamma_\mu u_t) \,,
&
\left[ Q_{dd} \right]_{prst}  &= (\bar{d}_p \gamma^\mu d_r)(\bar{d}_s \gamma_\mu d_t) \,,\\
\left[ Q^{(1)}_{qu} \right]_{prst} &= (\bar{q}_p \gamma^\mu q_r)(\bar{u}_s \gamma_\mu u_t) \,,
&
\left[ Q^{(1)}_{qd} \right]_{prst} &= (\bar{q}_p \gamma^\mu q_r)(\bar{d}_s \gamma_\mu d_t) \,,
\\
\left[ Q^{(8)}_{qu} \right]_{prst} &= (\bar{q}_p \gamma^\mu T^A q_r)(\bar{u}_s \gamma_\mu T^A u_t) \,,
&
\left[ Q^{(8)}_{qd} \right]_{prst} &= (\bar{q}_p \gamma^\mu T^A q_r)(\bar{d}_s \gamma_\mu T^A d_t) \,.
\end{align}

\paragraph*{Down-quark magnetic dipoles:}
The operators generating $b\to s\gamma$ after EW breaking are
\begin{equation}
Q_{dW}^{ij} = (\bar{q}_i \sigma^{\mu \nu} d_j) \tau^I H W^I_{\mu \nu}\,,
\qquad
Q_{dB}^{ij} = (\bar{q}_i \sigma^{\mu \nu} d_j) H B_{\mu \nu} \,.
\end{equation}

\paragraph*{Modified Higgs couplings:}
Here we have
\begin{equation}
Q_{uH}^{ij} = (H^\dag H)(\bar{q}_i u_j \tilde{H})\,,
\qquad
Q_{dH}^{ij} = (H^\dag H)(\bar{q}_i d_j {H})
\,.
\end{equation}

\paragraph*{The $W$ mass:}
The prediction for the $W$-boson mass is affected already at tree-level by
\begin{equation}
Q_{HD} = (H^\dag D_\mu H)^* (H^\dag D^\mu H) \,,
\qquad
Q_{HWB} = (H^\dag \tau^I H) W^I_{\mu \nu} B^{\mu \nu}\,.
\end{equation}

\section{CKM treatment}

Since we modify $W$-quark couplings at tree-level, the determination of the ``correct'' CKM elements is non-trivial. In \smelli, the method of Ref.~\cite{Descotes-Genon:2018foz} is implemented which fixes the four inputs needed to determine the CKM using four observables, fully taking into account both the SM and new physics contributions, at each point in parameter space. These four observables are $\Gamma(K^+ \to \mu^+ \nu) / \Gamma(\pi^+ \to \mu^+ \nu)$, $\mathrm{Br}(B \to X_c e^+ \nu)$, $\mathrm{Br}(B^+ \to \tau^+ \nu)$, and $\Delta M_d / \Delta M_s$.
Once the CKM matrix is fixed, it is then used as the input to all the other theory predictions, including the ``SM'' part. Note however that this means there is essentially a scheme dependence to which observables show discrepancies with data (since the four used to fix the CKM must agree with experiment, by construction), while only the global $\Delta \chi^2$ is physical.

\section{VLQ to SMEFT matching coefficients}

The part of the Lagrangian detailing the interaction between the SM Higgs and two VLQs is given by
\begin{equation}
\begin{aligned}
-\mathcal{L}^H_{\mathrm{VLQ VLQ}} =&
\lambda _{}^{L,UQ}{{\bar U} }{{\tilde H}^\dag } P_L {Q_{1} } + \lambda _{}^{L,DQ}{{\bar D} }{H^\dag } P_L {Q_{1} }
+ \lambda _{}^{L,{Q_5}D}{{\bar Q}_{5,\alpha }}\tilde H P_L {D_\beta } + \lambda _{}^{L,{Q_7}U}{{\bar Q}_{7 }}H P_L {U }
\\
&+ \frac{1}{2}\lambda _{}^{L,{T_1}Q}{H^\dag }\tau \cdot{{\bar T}_{1 }} P_L {Q_{1} }
+ \frac{1}{2}\lambda _{}^{L,{T_2}Q}{{\tilde H}^\dag }\tau \cdot{{\bar T}_{2 }} P_L {Q_{1} }
+ \frac{1}{2}\lambda _{}^{L,{T_1}{Q_5}}{\tilde H^\dag }\tau \cdot{{\bar T}_{1 }} P_L {Q_{5 }}
\\
&+ \frac{1}{2}\lambda _{}^{L, {T_2}{Q_7}}{{ H}^\dag }\tau \cdot{{\bar T}_{2 }} P_L {Q_{7 }}
+ (L \leftrightarrow R) + \hc \,.
\end{aligned}
\end{equation}
Note that, generalizing Ref.~\cite{deBlas:2017xtg}, the interaction between two VLQs and the SM Higgs can be different for the left-handed and right-handed components.

We give the coefficients relevant for our phenomenological study in the main text. At tree-level, in addition to those given in the main text (Eq.~(II.10)), , we also find for the Wilson coefficients of the three Higgs operator
\begin{align}
C_{uH}^{ij} &=
\frac{Y^u_{kj} \xi^U_{k} \xi^{U *}_{i}}{2 M_{U}^2}
+ \frac{Y^u_{ik} \xi^{u_1}_{j} \xi^{u_1*}_{k}}{2 M_{Q_{1 }}^2}
+ \frac{Y^u_{ik} \xi^{Q_7}_{j} \xi^{Q_7 *}_{k}}{2 M_{Q_{7}}^2}
\,.
\end{align}

At one-loop, we only present here the one-loop matching terms proportional to either the new physics coupling $\xi$, or the SM up-type Yukawa $Y^u$, thus neglecting the down-type and lepton Yukawas as well as the gauge couplings $g_1, g_2$.
One exception are the electroweak-dipole operators, where we include $Y^d g_{1,2} \xi^2$ terms since the SM dipole operators are already suppressed by the same factor.
Another exception is the $W$ mass operator $Q_{HWB}$ where we also include $g_1 g_2 \xi^2$ coefficients, since this operator can potentially have large contribution (see Eq.~(II.9)). In principle we also consider $\alpha_s$ terms but these only arise in a) the four-quark operators with a Kronecker delta, which therefore cannot induce a change of flavour, b) totally bosonic operators which are not interesting for our purposes, and c) gluon dipoles at one-loop, which are too small to generate an observable effect.

The main technicality associated with matching is to correctly account of the one-loop renormalization of the SMEFT, specifically the terms proportional to the SM quark Yukawas, as these allow the $Q_{Hq}^{(1,3)}$ operators to contribute to the renormalization of the $Q_{qq}^{(1,3)}$ operators, and similarly for the other $\Delta F=2$ operators. We perform the calculation by matching off-shell amplitudes calculated in the full VLQ theory and the SMEFT, and allowing for a $\overline{\rm MS}$ counter-term contribution on the SMEFT side.
A potential IR divergence is regulated using the Higgs doublet mass $\mu_H$, which is set to zero at the end of the calculation.

\paragraph*{One-loop matching for $\Delta F=2$ processes:}

\begin{figure}[t]
\includegraphics[width=0.7\textwidth]{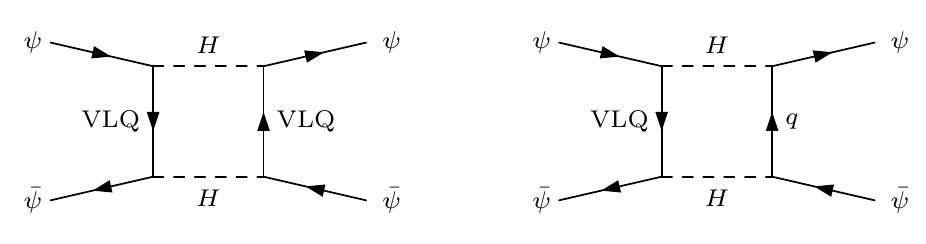}
\caption{Typical box diagrams giving rise to the $\Delta F=2$ matching coefficients, where the external fermions $\psi$ can be any of the SM quarks $\{q_L, u_R, d_R\}$, and in the right diagram the internal $q$ represents any SM quark and gives rise to the Yukawa corrections.
}
\label{fig:df2_matching_boxes}
\end{figure}

The one-loop matchings onto the $\Delta F=2$ operators
are obtained from the diagrams in Fig.~\ref{fig:df2_matching_boxes}.
Among the three representations ($U$, $Q_1$ with $\xi^{u_1}$, and $Q_7$), only $U$ contributes to \bbms.
The one-loop matching condition for $U$ is
\begin{align}
\left[ C^{(1)}_{qq} \right]_{ijij} &=
- \frac{\xi^{U \ast}_{i} \xi^{U}_{j}
\xi^{U \ast}_{ i} \xi^{U}_{ j}  }{256 \pi^2 M_U^2}
+ \frac{\xi^{U*}_i \xi^U_j (Y^u Y^{u \dagger})_{ij}}{128 \pi^2}  \widetilde{F}(M_U)
\,,
\\
\left[ C^{(3)}_{qq} \right]_{ijij}  &=
- \frac{\xi^{U \ast}_{ i} \xi^{U}_{ j} \xi^{U \ast}_{ i} \xi^{U}_{ j} }{256 \pi^2 M_U^2}
+ \frac{\xi^{U*}_i \xi^U_j (Y^u Y^{u \dagger})_{ij}}{128 \pi^2}  \widetilde{F}(M_U)
\,,
\end{align}
where the IR-finite loop function is
\begin{equation}
\widetilde{F}(m) = \frac{1}{m^2} \left( \frac{3}{2}+  \ln \frac{\mu^2}{m^2} \right)\,,
\end{equation}
and the renormalization scale $\mu$ should be $\mathcal{O}(M_{\rm VLQ})$.

\begin{figure}[t]
\includegraphics[width=0.7\textwidth]{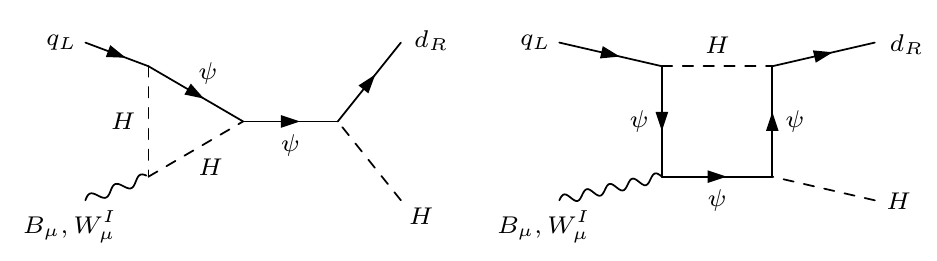}
\caption{Typical diagrams giving rise to the dipole matching coefficients, where the internal fermions $\psi$ can be a SM quark or a VLQ.
}
\label{fig:dipole_matching_examples}
\end{figure}

\paragraph*{One-loop matching for down-quark magnetic dipoles:}
Some partial work was done in Ref.~\cite{Crivellin:2021rbq} for the dipole operators, but note that we find our VLQ interactions are more general than those considered in that work, and additional diagrams contribute to the one-loop matching onto the dipole operators $C_{dB}$ and $C_{dW}$. Some typical diagrams are shown in \cref{fig:dipole_matching_examples}.
Among the three representations,
only $U$ produces the one-loop matching condition,
\begin{align}
\left[ C_{dB} \right]_{ij} &=
 \frac{7 g_1 Y^d_{kj} \xi^{U*}_i \xi^U_k}{1152 \pi^2 M_U^2}
\,, \qquad
\left[ C_{dW} \right]_{ij} =
- \frac{5 g_2 Y^d_{kj} \xi^{U*}_i \xi^U_k}{384 \pi^2 M_U^2}
\,.
\end{align}

\begin{table}[t]
\newcolumntype{C}{>{\centering\arraybackslash}p{20px}}
\centering
\renewcommand{\arraystretch}{1.2}
\begin{tabular}{lCCCCCCC}
\toprule
Tree & $U$ & $D$ & $Q_1$ & $Q_5$ & $Q_7$ & $T_1$ & $T_2$ \\
\midrule
$Z$-$u_L^i$-$u_L^j$ & \checkmark & $\times$ & $\times$ & $\times$ & $\times$ & \checkmark & \checkmark \\
$Z$-$u_R^i$-$u_R^j$ & $\times$ & $\times$ & \checkmark & $\times$ & \checkmark & $\times$ & $\times$ \\
$Z$-$d_L^i$-$d_L^j$ & $\times$ & \checkmark & $\times$ & $\times$ & $\times$ & \checkmark & \checkmark \\
$Z$-$d_R^i$-$d_R^j$ & $\times$ & $\times$ & \checkmark & \checkmark & $\times$ & $\times$ & $\times$ \\
\bottomrule
\end{tabular}
\caption{Overview on modified $Z$-quark couplings (in broken $SU(2)_L$) at tree level in the  VLQ models.}
\label{tab:Zqqtable}
\end{table}

\paragraph*{One-loop matching for modified gauge boson couplings:}
While the tree-level SMEFT operators, generated by the VLQs, can affect $W$ and $Z$ couplings, the low-energy $Z$ coupling to up- and down-type quarks specifically depends on $C_{Hq}^{(1)} - C_{Hq}^{(3)}$, $C_{Hu}$ and $C_{Hq}^{(1)} + C_{Hq}^{(3)}$, $C_{Hd}$, respectively (see Eq.~(II.10)). As we can see that this leads to some of the $Z$ quark couplings remaining SM-like for certain VLQs (as summarised in \cref{tab:Zqqtable}). Since these interactions are constrained by EWPO, and also contribute to many interesting processes such as $b \to s \ell \ell$, we also calculate the one-loop matching for the $U, Q_1, Q_7$ cases where they are not already present at tree level.
\begin{align}
\left[ C_{Hq}^{(1)} \right]_{ij} =&
 {
\frac{\xi_j^U \xi_k^{U *} Y_{il}^u Y_{lk}^{u \dagger}}{32 \pi ^2 M_U^2} \left(1 + \ln \frac{\mu^2}{M_U^2} \right)
}
 {
+\frac{\xi_k^U  \xi_i^{* U} Y_{kl}^u Y_{lj}^{u \dagger}}{32 \pi ^2 M_U^2} \left(1 + \ln \frac{\mu ^2}{M_U^2} \right)
}
{
-\frac{\xi _j^U \xi _k^U \xi _i^{* U} \xi _k^{* U}}{256 \pi ^2 M_U^2} \left( 17 + 14 \log \frac{\mu ^2}{M_U^2} \right)
}
\notag \\
&
-\frac{Y^u_{il} Y^{u \dagger}_{kj} \xi^{u_1}_k \xi^{u_1 *}_l}{384 \pi^2 M_{Q_1}^2} \left( 1 + 6\ln \frac{\mu^2}{M_{Q_1}^2} \right)
{
-\frac{Y^u_{il} Y^{u \dagger}_{kj} \xi^{Q_7}_k \xi^{Q_7 *}_l}{64 \pi^2 M_{Q_7}^2} \left( -13 + 6\ln \frac{\mu^2}{M_{Q_7}^2} \right)
}
\,, \\
\left[ C_{Hq}^{(3)} \right]_{ij} =&
\frac{\xi_j^U \xi_k^U \xi_i^{U *} \xi_k^{U *}}{256 \pi^2 M_U^2} \left( 9 + 14\ln \frac{\mu ^2}{M_U^2} \right)
-\frac{\xi^{u_1}_k \xi^{u_1 *}_l Y^u_{il} Y^{u \dagger}_{kj}}{96 \pi^2 M_{Q_1}^2}
-\frac{5 \xi^{Q_7}_k \xi^{Q_7 *}_l Y^u_{il} Y^{u \dagger}_{kj} }{192 \pi^2 M_{Q_7}^2}
\,.
\end{align}

\paragraph*{One-loop matching for the $W$ mass:}
\begin{figure}[t]
\includegraphics[width=0.6\textwidth]{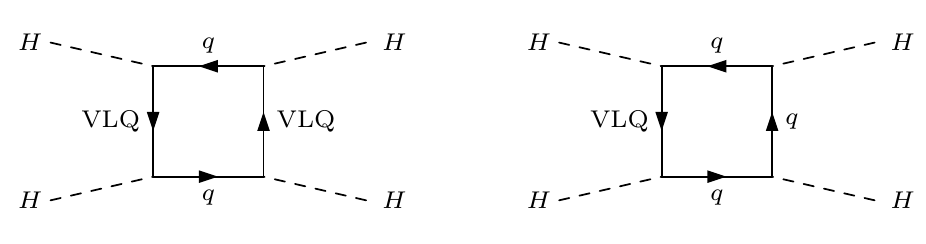}
\caption{Typical box diagrams giving rise to the $W$ mass,
where the internal fermions $q$ represents any SM quark.
}
\label{fig:QHD_matching_boxes}
\end{figure}
The one-loop matching onto the operators $Q_{HD}$ and $Q_{HWB}$, which modify the $W$-boson mass prediction, are obtained from the diagrams in Fig.~\ref{fig:QHD_matching_boxes}.
The matching conditions for the three representation VLQs are
\begin{align}
C_{HD} =&-  \frac{ 3\xi^U_i \xi^U_j  \xi^{U\ast}_i \xi^{U\ast}_j }{32\pi^2 M_U^2}
+ \frac{3\xi^U_i (Y^u Y^{u \dagger})_{ij} \xi^{U *}_j}{16 \pi^2} F_1(M_U)
\nonumber
\\
&-  \frac{ \xi^{u_1}_i \xi^{u_1}_j  \xi^{u_1\ast}_i \xi^{u_1\ast}_j }{8\pi^2 M_{Q_1}^2}
+ \frac{3\xi^{u_1}_i (Y^{u \dagger} Y^u )_{ij} \xi^{u_1 *}_j}{8 \pi^2} F_2(M_{Q_1})
-  \frac{ \xi^{Q_7}_i \xi^{Q_7}_j  \xi^{{Q_7}\ast}_i \xi^{{Q_7}\ast}_j }{8\pi^2 M_{Q_7}^2}
- \frac{3\xi^{Q_7}_i (Y^{u \dagger} Y^u )_{ij} \xi^{Q_7 *}_j}{8 \pi^2} F_2(M_{Q_7})
\,,
\\
C_{HWB} &=
- \frac{g_1 g_2 \xi^U_i \xi^{U\ast}_i }{64\pi^2 M_U^2}
-  \frac{g_1 g_2 \xi^{u_1}_i \xi^{u_1\ast}_i }{96\pi^2 M_{Q_1}^2}
% -  \frac{g_1 g_2 \xi^d_i \xi^{d\ast}_i }{192\pi^2 M_{Q_1}^2}
+  \frac{g_1 g_2 \xi^{Q_7}_i \xi^{{Q_7}\ast}_i }{96 \pi^2 M_{Q_7}^2} \,,
\end{align}
with
\begin{equation}
F_1 (m) = \frac{1}{m^2} \left( \frac{1}{2} + \ln \frac{\mu^2}{m^2} \right) \,,\quad
F_2 (m) = \frac{1}{m^2} \left( \frac{3}{2} + \ln \frac{\mu^2}{m^2} \right) \,.
\end{equation}

\section{$t \to c Z$ and $t \to c h$}

For $t\to cZ$~\cite{Liu:2020bem}, we take the branching ratio to the $Z$ boson to be:
\begin{equation}
\mathrm{Br} (t \to c Z) \approx 0.47 \left|\lambda_{tcZ}\right|^2 \,,
\end{equation}
with the couplings defined by the Lagrangian terms
\begin{equation}
\mathcal{L} = \frac{g}{2 c_W} \lambda_{tcZ} Z_\mu \bar{c} \gamma^\mu (\lambda_L P_L + \lambda_R P_R) t + \mathrm{h.c.}\,,
\end{equation}
with the normalisation \(\lambda_L^2 + \lambda_R^2 = 1\).
In Ref.~\cite{ATLAS:2018jqi}, they give the simple formula for the FCNC Higgs branching ratio:
\begin{equation}
\mathrm{Br} (t \to c h) \approx 0.27 \left|\lambda_{t_L c_R h}^2 + \lambda_{c_L t_R h}^2 \right| \,,
\end{equation}
where the couplings are defined by the Lagrangian terms
\begin{equation}
\mathcal{L} = \lambda_{t_L c_R h} \bar{t} P_R c \, h + \lambda_{c_L t_R h} \bar{c} P_R t \, h + \mathrm{h.c. } \,.
\end{equation}
Comparing to the SMEFT Lagrangian, we see the correspondence:
\begin{align}
\lambda_{tcZ} \lambda_L &= v^2 \left[ V \cdot \left( C_{Hq}^{(3)} - C_{Hq}^{(1)} \right) V^\dagger \right]_{23} \,,
\quad
\lambda_{tcZ} \lambda_R = v^2 \left[ C_{Hu} \right]_{23} \,,
\\
\lambda_{t_L c_R h} &= -\frac{v^2}{\sqrt{2}} \left[ C_{uH} \right]_{32} \,,
\quad
\lambda_{c_L t_R h} = -\frac{v^2}{\sqrt{2}} \left[ C_{uH} \right]_{23} \,.
\end{align}
For the estimation of the $t \to c Z,h$ branching ratios, at the level of precision we are considering, we can take the CKM matrix to be the unit matrix.

\end{document}